\documentclass[12pt]{iopart}
\usepackage{iopams}  

\usepackage{graphicx}
\usepackage{dcolumn}
\usepackage{bm}

\usepackage{graphicx}

\begin{document}

\title{Unhappy Vertices in Artificial Spin Ice: \\ New Degeneracies from Vertex-Frustration.}
\author{Muir J. Morrison, Tammie R. Nelson, and Cristiano Nisoli}

\address{Theoretical Division and Center for Nonlinear Studies, Los Alamos National Laboratory, Los Alamos, NM 87545}

\ead{cristiano@lanl.gov, cristiano.nisoli@gmail.com}

\date{\today}

\begin{abstract}
In 1935, Pauling estimated the residual entropy of water ice with remarkable accuracy by considering the degeneracy of the ice rule {\it solely at the vertex level}. Indeed, his estimate works well for both the three-dimensional pyrochlore lattice and the two-dimensional six-vertex model, solved by Lieb in 1967.
The case of honeycomb artificial spin ice is similar: its pseudo-ice rule,  like the ice rule in Pauling and Lieb's systems, simply extends  a degeneracy which is already present in the vertices to the global ground state. The anisotropy  of the  magnetic interaction limits the design of  inherently degenerate vertices in artificial spin ice, and the honeycomb  is the only degenerate array produced so far. In this paper we show how to engineer  artificial spin ice in a virtually infinite variety of  degenerate  geometries built out of  non-degenerate vertices. In this new class of vertex models, the residual entropy  follows not from a freedom of choice at the vertex level, but from the nontrivial relative arrangement of the vertices themselves. In such arrays, loops exist along which not all of the vertices can be chosen in their lowest energy configuration: these loops are therefore vertex-frustrated since they contain unhappy vertices. Residual entropy emerges in these lattices as configurational freedom in allocating the unhappy vertices of the ground state.  These  new geometries will finally allow for the fabrication of many novel extensively degenerate artificial spin ice. 
\end{abstract}


\maketitle

\section{Introduction}

In the past century, the application of concepts from physics to material sciences has brought a profound understanding of  naturally occurring materials. As the discipline moves from descriptive science to design science,   theoretical ideas, especially  on collective behaviors,  are exploited in the design of metamaterials with desired emergent properties.

Artificial Spin Ice (ASI) is a perfect example of this trend. ASI is a two dimensional  array of magnetic nano-islands which can be engineered to desired geometries. It owes its name to  naturally occurring magnetic materials, the ``spin ice'' titanates~\cite{Ramirez1999, Bramwell2001}, as well as to the various ice-type models of statistical mechanics which it was designed to mimic.  It was introduced as a system in which frustration can be manipulated and experimentally tuned, in contrast to naturally occurring spin ice~\cite{Wang2006}. ASI also permits direct imaging of individual microstates, thus providing a powerful test of statistical mechanical tools and concepts. While this material-by-design approach addresses fundamentally relevant technological issues connected to the high density storage limit in industrial applications, it has also opened a new arena to study collective phenomena in artificial systems whose degrees of freedom can be tailored. 

Early realizations  have shown  that  many of ASI's features can often be approximated as a vertex model~\cite{Wang2006, Nisoli2007, Nisoli2010}. Two-dimensional vertex models of statistical mechanics were introduced in the 60's to study  the zero temperature entropy in water ice and have since evolved  into an independent field of mathematical physics (See~\cite{Baxter1981} and references therein). These models are generally based on a (square) lattice in which Ising spins are allocated on the edges, and energies are assigned to different vertex configurations--hence the name, vertex models. In the six-vertex models for instance, only vertices obeying the ice rule are allowed. 

\begin{figure}
\includegraphics*[scale=0.47]{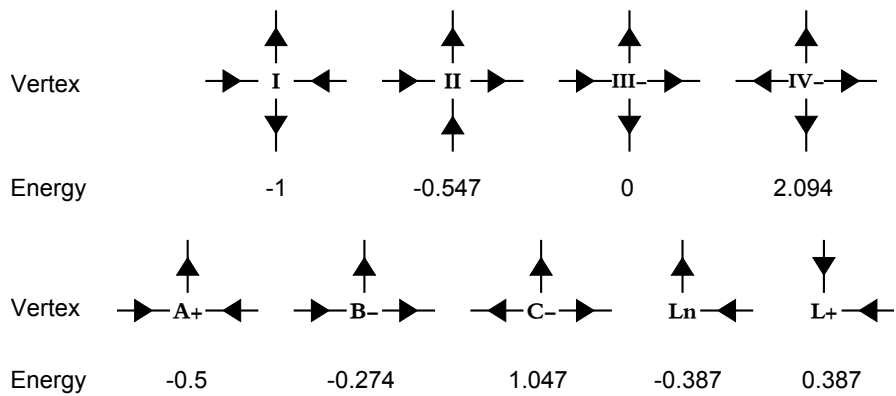}
\caption{
Top: the four-legged vertices that are used in square ASI. Type I and Type II obey the ice rule, but are not degenerate: because of the anisotropy of the magnetic interaction, perpendicular spins interact more strongly than collinear spins. Bottom: the three-legged vertices used in brickwork lattice. Type A and B obey the pseudo-ice rule, yet the anisotropy lifts the degeneracy. Also shown are two-legged vertices which intervene in some of our lattices.  Below each vertex we report a choice for   the energy  which is described in the text. 
Note that, within the dumbbell model, Types I, II, and Ln are uncharged, while Types III, IV, A, B, C, and L+ all carry a magnetic monopole, hence the \(+/-\) to distinguish the charge.
 \label{Fig:VertEn}
}
\end{figure}

\begin{figure}
\includegraphics*[scale=0.32]{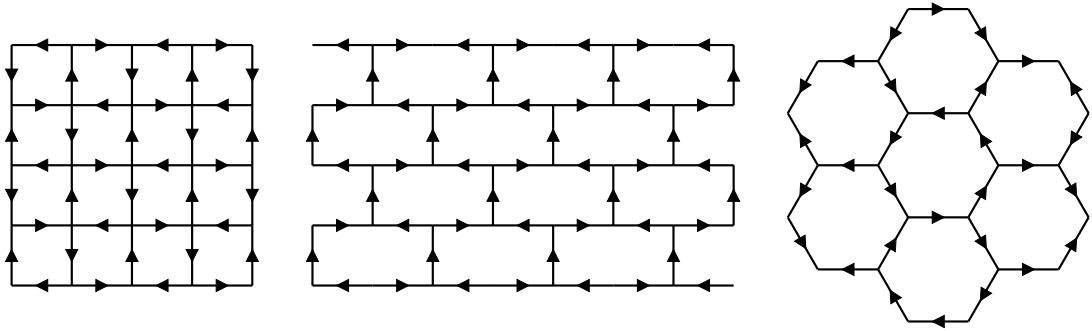}
\caption{
Square, brickwork, and hexagonal ASI. Lines represent nanomagnet islands, with each arrow indicating the orientation of its island's magnetic moment in their vertex  ground state. While square and brickwork ASI have a unique ground state, hexagonal ASI exhibits residual entropy.  
 \label{Fig:SqLatt}
 }
\end{figure}

When ASI is modeled as a vertex system, the energetic hierarchy of its vertices is inherited  from the magnetic interaction among  nanoislands. Unfortunately,  realistic magnetic interactions  limit the possibility of reproducing a specific vertex model--although nanofabrication techniques have been proposed to circumvent this aspect~\cite{Moller2006}. For instance,  differences in the pairwise island interaction energies for neighbors at $\pi $ and $\pi/2$ angles lifts the vertex degeneracy of the ice rule (and pseudo-ice rule) for perpendicular four-legged (and three-legged) vertices in square and brickwork ASI (Fig.~\ref{Fig:VertEn}). This lack of vertex degeneracy (beyond the obvious spin inversion) leads  to a well defined antiferromagnetic ground state for square ASI (Fig~\ref{Fig:SqLatt}). Therefore, square  ASI is described at the vertex level by a generalized F-model rather than an extensively degenerate six-vertex model~\cite{Baxter1981,Wu1969}. The lack of degeneracy does not detract from its interest, and square ASI has raised many new distinct issues at the confluence of thermodynamics and granular materials, at the  interplay between macroscopic demagnetization and microscopic energetics, and as a possible medium to study magnetic monopoles unbinding~\cite{Wang2006, Nisoli2007, Nisoli2010, Ke2008, Mol2009, Morgan2010,Nisoli2012, Kapaklis2012,Greaves2012}. Similarly, the ladder or brickwork lattice (Fig.~\ref{Fig:SqLatt}), topologically equivalent to the hexagonal lattice, possesses a defined ground state when modeled as a vertex system~\cite{Li2010}.

On  the other hand, honeycomb ASI (Fig.~\ref{Fig:SqLatt})---in which the nanoislands are placed on the edges of a honeycomb lattice, and therefore on the vertices of a Kagome lattice---exhibits a genuine degeneracy in the energy hierarchy of the vertices:  the six vertices obeying the pseudo-ice rule (one-in/two-out, two-in/one-out) possess the lowest energy~\cite{Qi2008}. Its ground state extensively inherits this degeneracy  which leads  to a zero temperature entropy. The latter has been extracted directly~\cite{Lammert2010} \footnote{Of course this is only true  in the vertex approximation. Although signatures of long range dipolar effects have been reported~\cite{Rougemaille2011}, inclusion of long range interactions should reveal new and intriguing phases~\cite{Moller2009,Chern2011} which have not yet been experimentally observed}.

As of today, hexagonal ASI represents the only extensively degenerate version of ASI because  the anisotropy of the magnetic interaction makes it difficult to produce degenerate energetics at the vertex level for other geometries. Here we propose to circumvent the problem of ASI residual entropy by showing that  it is possible to design lattices  in which the extensive degeneracy does not arise from built-in vertex degeneracy, but rather from the mutual arrangement of non-degenerate vertices. This, we think, will open new directions in ASI design: not only because it can lead to a virtually infinite variety of degenerate ASI, but also because vertex-frustration can lead to custom design of relatively mobile topologically protected excitations, which might be field-driven.

\section{Vertex frustration  {\it vs.} local pairwise frustration}
\label{Sec:Global Frustration}

In statistical mechanics, a vertex system is  a lattice of spins of variable geometry in which  energies (and therefore Boltzmann weights) are assigned to vertex configurations. Notable examples are the six-vertex model, the F-model, the KDP model, and the eight vertex model~\cite{Baxter1981}. The ground state of a vertex model corresponds to spin arrangements that minimize the overall vertex energy and  might or might not possess residual entropy. When it does, the degeneracy of the ground state follows from a degeneracy in the vertex energetics. In fact, Pauling was able to estimate the residual entropy of water-ice by only reasoning in terms of vertex degeneracy.  

In physical realizations, this local degeneracy is often the consequence of frustrated  interactions. For instance, local geometric frustration in ASI arises at the vertex level  when islands are arranged  such that pairwise interactions cannot be simultaneously minimized. This frustration might or might not lead to an extensive degeneracy of the global ground state. In current  realizations of ASI,  the global degeneracy of the ground state is  a consequence of the local  degeneracy at the vertex level,  which in turn is the consequence of  frustration in pairwise interactions. As in real life, frustration does not guarantee freedom of choice, although choice is born from frustration. Pairwise frustration might present multiple equally expensive choices of vertices, and  the degeneracy at the vertex level is then inherited by the extended lattice. 

From now on we will neglect this local form of frustration of pairwise interactions, whose only  effect is the assignment of the energy hierarchy to the vertices.  Instead we will  look for a broader kind of frustration as a different source of choice/degeneracy. 

\begin{figure}
\includegraphics*[scale=0.28]{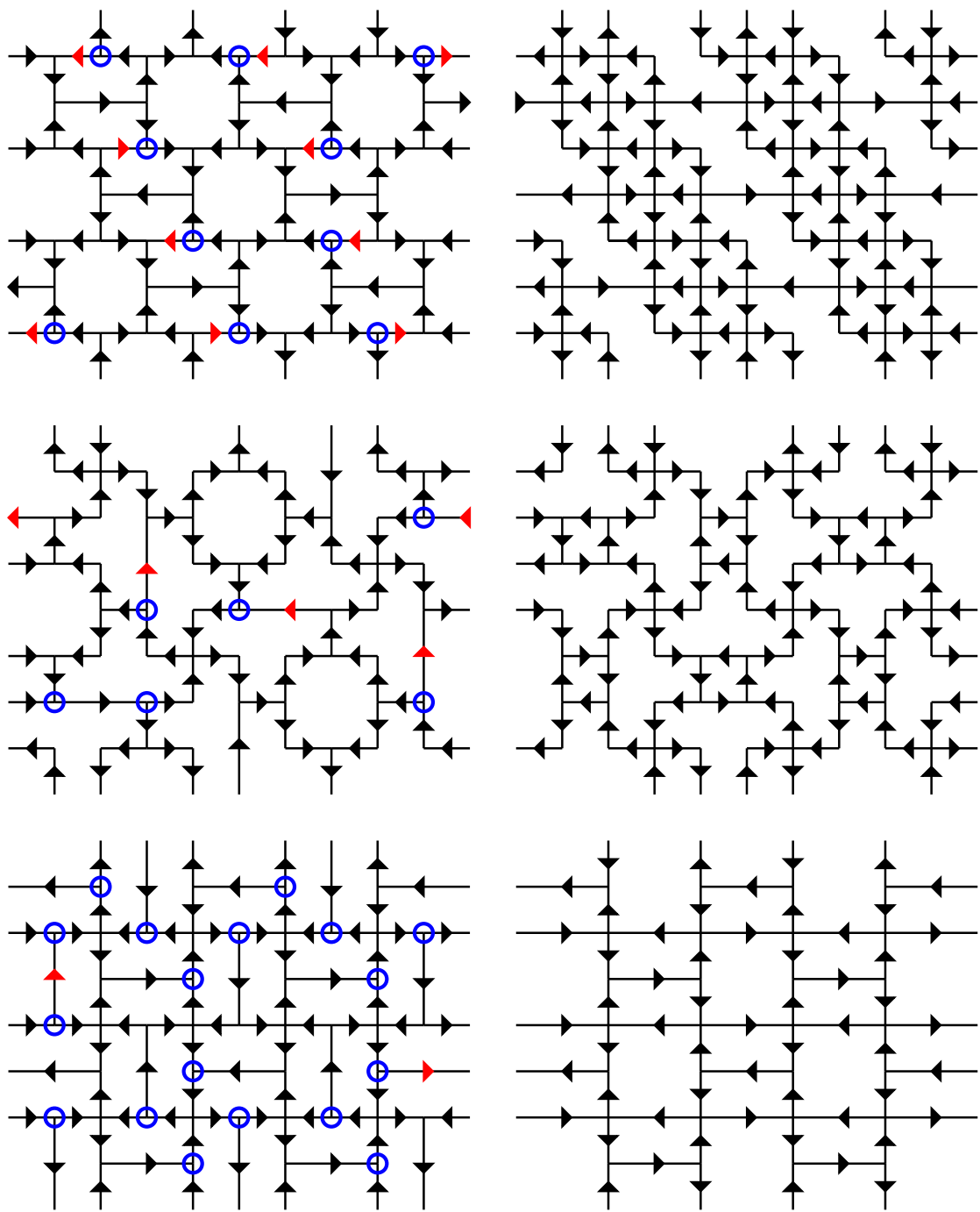}
\caption{
(Color online)
Vertex-frustrated and nonfrustrated lattices based on perpendicular vertices. Lattices on the left are frustrated while those on the right, despite the apparent similarity of the lattices, are not. The unhappy vertices are circled (blue), while red spins may be chosen in either direction without changing the energy of the configuration: that is not the only source of degeneracy, as other allocations of the unhappy vertices are degenerate. We will show that the 
``staggered brickwork'' lattice (top left) has a ``trivial'' degeneracy, which can be mapped into a system of independent spins. In contrast, the ``pinwheel'' (middle left) and ``shakti'' lattices (lower left) possess more complex degeneracy.
\label{Fig:TopFr}
 }
\end{figure}

\subsection{Vertex-Frustration}

We will say that a vertex model is \textit{vertex-frustrated}, when it is impossible to find a global arrangement of spins such that all the vertices are in  their lowest energy configuration.  Equivalently, the ground state of a vertex-frustrated model must contain excited vertices. 

Note that these excited vertices are {\it not} excitations of the global system. They are ``excited'' (by the topology of the lattice) because, as vertices, their energy is higher than the minimum energy for a vertex.  In a sense, they are topologically protected excitations which cannot be eliminated from the ground state. We will refer to these as ``unhappy vertices'' to avoid confusion and distinguish them from the elementary excitations \textit{above} the ground state. We call this frustration vertex-frustration  because it arises from the frustrated attempt to allocate each vertex of the array in its minimum energy configuration, rather than from the local pairwise interaction between spins. From now on, by {\it frustration} we will always be referring to {\it  vertex-frustration}. 

Clearly this frustration, as with any frustration, has the potential to generate extensive degeneracy insofar as the number of choices in allocating those topologically protected unhappy vertices grows exponentially with the size of the system. 

\subsection{Vertex-frustration of non-degenerate vertices}

In general, vertex-frustrated models can be considered independently of the fact that the vertex energetics is degenerate, nondegenerate, or mixed. However, keeping in mind the physical realization of these lattices as novel degenerate ASI, we will work with systems of perpendicular vertices: more specifically, the vertices reported in Fig.~\ref{Fig:VertEn} which have only two configurations of lowest energy, identical up to spin flip symmetry.  By using only square and brickwork style vertices, we eliminate degeneracy arising from local geometric frustration. These vertices can be easily nano-fabricated, and were in fact the first vertices to be presented in ASI~\cite{Wang2006, Li2010a} .

Figure~\ref{Fig:TopFr} provides examples of vertex-frustrated and non-frustrated lattices, built with vertices of 2, 3, and 4 legs, and based on the vertex energetics of Fig.~\ref{Fig:VertEn}. The reader could try to eliminate the unhappy vertices (blue circles) in the lattices on the left, only to realize that any collective spin switch that removes unhappy vertices creates as many of them, or more~\footnote{There is a subtlety here: with appropriate spin flips, it is possible to reduce the \textit{number} of unhappy vertices by trading Type Bs for Type IVs, but the total \textit{energy} of such a configuration would be higher.}. 
The lattices on the right of Fig~\ref{Fig:TopFr} are presented in their ground state. Figure~\ref{Fig:TopFr} raises a series of questions: can we decide when a lattice is vertex-frustrated without solving for its ground state? Are all vertex-frustrated lattices degenerate? Is vertex-frustration the only way to engineer  residual entropy with non-degenerate vertices? Are there general theoretical tools we can employ to describe the ground state of a vertex-frustrated lattice, and to compute its entropy?

\begin{figure} 
\includegraphics*[scale=0.22]{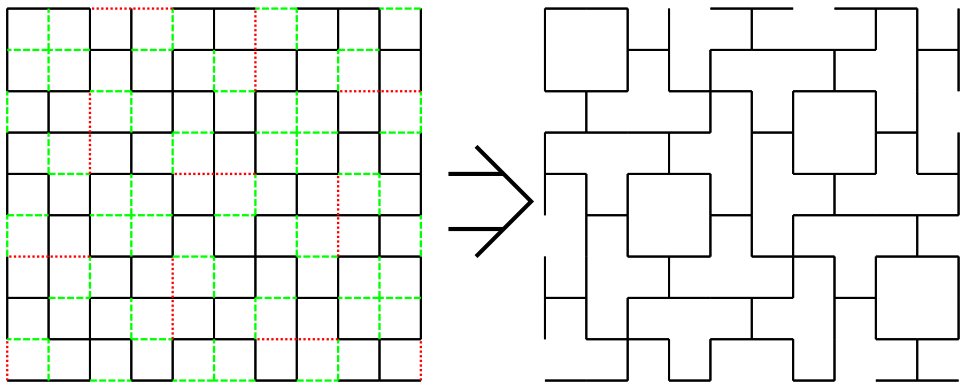}
\caption{
(Color online)
Islands are removed from a square lattice (dashed, green) to form a  frustrated pinwheel lattice. In particular, if two islands opposite a single vertex are removed (dotted, red), the remaining two islands are fused into a single island.
 \label{Fig:Fuse}
}
\end{figure}

We start by noticing that just as the brickwork lattice can be constructed from a square lattice by removing islands, we can imagine constructing vertex-frustrated lattices of non-degenerate vertices by eliminating islands from a square lattice, with an extra condition, illustrated in Figure~\ref{Fig:Fuse}: when two opposite islands in a four-island vertex are removed, the remaining two islands must be fused to form a single long island. 
 
 In the case of a brickwork lattice, this decimation procedure does not involve fusion, and therefore the unique ground state can be inherited from the square lattice ground state. The presence of fusions is thus necessary, yet not sufficient, to grant both extensive degeneracy and global frustration to the new lattice:  at each fusion, a choice of direction must be made between the opposite pointing spins to be fused together. 
 
Fusion is clearly not a sufficient condition for frustration or degeneracy: the  square lattice itself, which is not vertex-frustrated nor degenerate,  can be deduced by decimation and fusion of  a larger square lattice. One might wonder, however, if nonfrustrated lattices can be topologically deformed into lattices in which all islands are the same length. For instance, the shakti's cousin in Figure~\ref{Fig:TopFr} (lower right) can be contracted horizontally so that all horizontal islands are the same length as the vertical islands. Yet even this is not always possible, as demonstrated by the nonfrustrated analog of the staggered brickwork (upper right in Figure~\ref{Fig:TopFr}): the long islands cannot be halved without creating new intersections, fundamentally changing the topology of the lattice.

One might expect that the decimation algorithm, by quantifying the number of choices made at each  fusion, could help compute the entropy of the ground state for the new frustrated lattice. This is true in some simple cases, which are therefore only trivially degenerate: their entropy is simply $S=N_l\ln2$ where $N_l$ is the number of long islands, which are free to flip (Fig.~\ref{Fig:TopFr},~\ref{Fig:Fuse}).  However this is not true in the general case, as  a look at the shakti lattice in Fig.~\ref{Fig:TopFr} can prove. To complicate the matter further: while it is true that one can obtain  vertex frustrated lattices of ``trivial'' degeneracy through decimation and fusion (see Appendix 1), it is not generally true that the ground state of a trivially degenerate  vertex-frustrated lattice can be obtained by decimation and fusion: an example is the staggered brickwork of Fig.~\ref{Fig:TopFr}.

\subsection{Energetics}

Note that repeated fusions in the decimation process  can introduce islands of many  different lengths. Because we are concerned with reasonable physical implementations in terms of ASI, we  limit ourselves to lattices with islands of \textit{two} different lengths. 
We would like the energy of a vertex to depend only on the configuration of spins and not on the length of the islands composing the vertex. This should be realizable by varying the aspect ratio or shape of the long islands compared to the short islands. However, as shown in the next section, most of the lattices that we have analyzed experience no change in the ground state as we relax our assumptions and assign a higher energy to  interactions involving longer islands.

Following previous work in real spin ice~\cite{Castelnovo2008} and artificial spin ice~\cite{Nisoli2010}, we treat each dipolar island as a "dumbbell" connecting pairs of oppositely charged monopoles. Figure~\ref{Fig:VertEn} tabulates the energies for all possible vertices in the lattices we consider. 
Note that, although we set
$(E_{\mathrm{II}}-E_{\mathrm{I}})/(E_{\mathrm{III}}-E_{\mathrm{I}}) = (\sqrt{2}-1)/(\sqrt{2}-1/2)$
as in~\cite{Nisoli2010}, we choose \(E_{\mathrm{III}}=0\) instead of \(E_{\mathrm{I}}=0\)).  See the Appendix for more detail on different energy modeling.

\begin{figure}
\includegraphics*[scale=0.25]{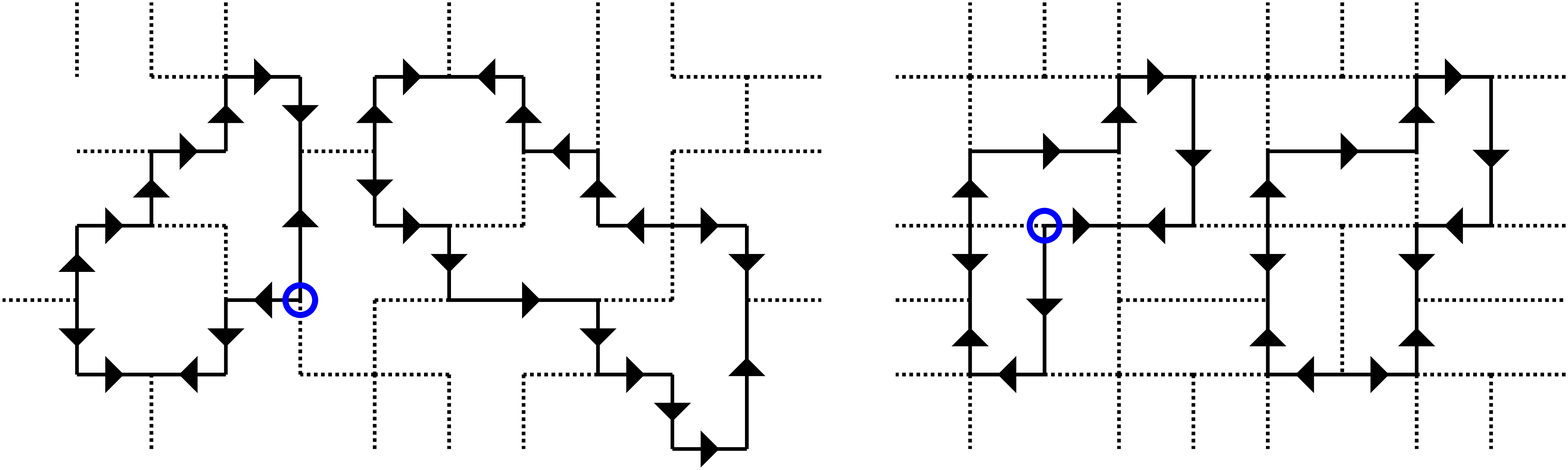}
\caption{
(Color online)
Loops in vertex-frustrated lattices. Solid lines denote the loops of interest while dotted lines indicate the background lattice. Spins are assigned beginning from an arbitrary spin choice anywhere on the loop. Upon ``crossing'' an intersection, the spin direction inverts, but when ``turning the corner'' the direction does not invert. 

Circles (blue) indicate the frustrated loops' failure to rejoin in a self-consistent manner (this could occur anywhere on the loop depending on one's choice of starting location). Even in a lattice such as the shakti, in which every minimal loop is frustrated, there exist nonminimal loops which are not frustrated.
 \label{Fig:spinLoops}
 }

\end{figure}

\section{Criteria for vertex-frustration}

It is often not clear what distinguishes frustrated from non-frustrated lattices by casual inspection of their geometry and we will now characterize  vertex-frustration.

\subsection{Loops}

A  \textit{ loop} is a continuous, closed chain of islands. A {\it minimal loop} is a loop that does not contain vertices in its interior. We  say that {\it a loop is vertex-frustrated} if it is impossible to choose every vertex of the loop in its lowest energy configuration. Clearly, a lattice is vertex-frustrated if and only if there is at least one  (and therefore infinitely many) vertex-frustrated loop(s).

In systems of non-degenerate vertices, like the ones we are considering here, the choice of any one spin in a vertex selects one of the two lowest energy configurations for the vertex (Fig.~\ref{Fig:VertEn}). Consequently, when dealing with frustrated loops, one can disregard vertices sitting on the loop and be concerned only with spin assignments  on the loop. Consider now  the following rule: assign spins on a loop such that  {\it the spins  invert their direction only when the loop has an intersection} (see Figure~\ref{Fig:spinLoops}). Then the loop is vertex-frustrated if and only if this  rule is violated. It follows that a loop is vertex-frustrated if and only if it has an odd number of intersections.

\begin{figure}
\includegraphics*[scale=0.34]{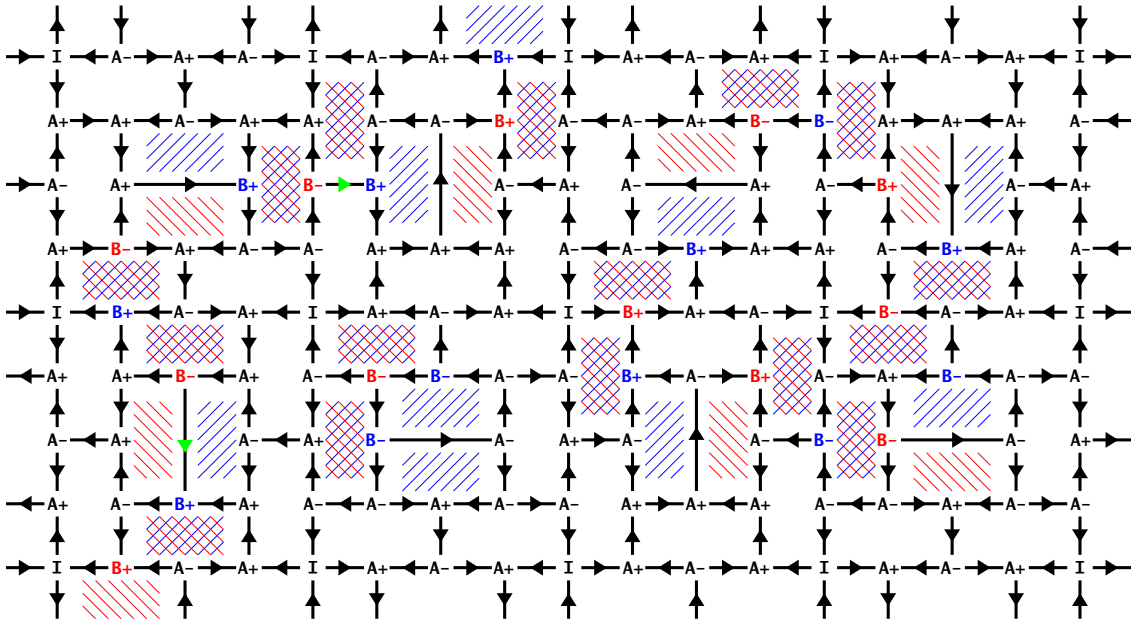}
\caption{
(Color online)
The Santa Fe lattice has a disordered ground state with a large residual entropy. As in Figure~\ref{Fig:defLoops}, hatching color corresponds to the color of the adjacent unhappy vertex providing the frustration. Note that many nonfrustrated loops are forced to contain unhappy vertices because of afferent frustrated loops; these loops are cross-hatched red and blue. The many possible ways of sharing unhappy vertices between neighboring loops makes the residual entropy  nontrivial to compute.
 \label{Fig:Tammies}
 }
\end{figure}

This reasoning can be extended to minimal loops. Indeed, there is at least one  vertex-frustrated loop if and only if there is at least a vertex-frustrated {\it minimal} loop. Consider a loop, and consider all the   minimal loops who are interior to it and afferent to it, as in Fig.~\ref{Fig:spinLoops}. With our perpendicular vertices in Fig.~\ref{Fig:VertEn},
 it is easy to prove that if none of these  minimal loops  are  frustrated, then the original loop is also not frustrated. Indeed, if the unhappy vertex of the original loop sits on a corner, it will frustrate the minimal loop sharing that corner; if it sits  on an intersection it will frustrate at least one of the two minimal loops sharing that intersection.
 
We will let the reader prove that this statement holds more generally:  {\it a lattice made of non-degenerate vertices with spin inversion symmetry is vertex-frustrated if and only if it contains some minimal loops with an odd number of intersections.} 

Even simpler is the characterization of frustrated loops in the lattices we are considering here, with  islands of only two different lengths,  the kind that can be deduced from decimation of a square lattice with single fusions. Then a loop has an odd number of intersections if and only if it has an odd number of long islands. In fact  the same loop in the original square lattice, before decimation/fusion,  possesses an even number of intersections. Each fusion removes an intersection. Therefore these  lattices are vertex-frustrated if and only if they contain minimal loops with an odd number of long islands. More generally, one can prove that lattices with islands of more than two lengths--which we do not consider here--are vertex-frustrated if and only if they contain minimal loops with long islands whose cumulative  length is odd (measured in units of the shorter island).

\subsection{Frustration by induction}
\label{Sec:FrustInduc}

A lattice is said to be {\it maximally frustrated} if every {\it minimal} loop is vertex-frustrated. That is the case of the shakti lattice, of which we will report more extensively elsewhere [Chern, Morrison and Nisoli, to be submitted]. The  staggered brickwork lattice on the other hand  is not maximally frustrated, and is trivially degenerate. The triviality of its degeneracy is not characteristic. 
Indeed, one can think of  lattices which are not maximally frustrated yet exhibit a  nontrivial ground state.  In these lattices the frustrated loops are afferent to non-frustrated ones and must induce an excited vertex on them. Therefore, nonfrustrated loops can become {\it frustrated by induction}. The many possible ways to achieve this induction make their ground state quite nontrivial to compute. An example is the Santa Fe lattice of Fig.~\ref{Fig:Tammies}.

\subsection{Frustration and degeneracy}

 Not only a  vertex-frustrated  system might be trivially degenerate, it might not be degenerate at all, and a counterexample of a vertex-frustrated lattice with no degeneracy is offered in Fig.~\ref{Fig:nonDegenFrust} of Appendix 1. Conversely, we know that the square and brickwork ASI are not vertex-frustrated and have no residual entropy. This is true in general: {\it if a vertex system  made of non-degenerate vertices  is not  vertex-frustrated then it  has no residual entropy}.  Indeed, assigning just one spin on the lattice selects the ground state of the vertex to which the spin belongs, which in turn selects the ground state of  all the neighboring vertices, and so on, covering the entire lattice. If that could not be done coherently, then there would be a vertex-frustrated loop, and therefore the lattice would be vertex-frustrated. Since this process can be made with only two  choices on the initial spin, it follows that the ground state has a degeneracy of two. 

Therefore  {\it in a system of non-degenerate vertices, vertex-frustration is the only way to achieve extensive degeneracy}.

\section{Arrays}
\label{Sec:Static}

\begin{figure}
\includegraphics*[scale=0.35]{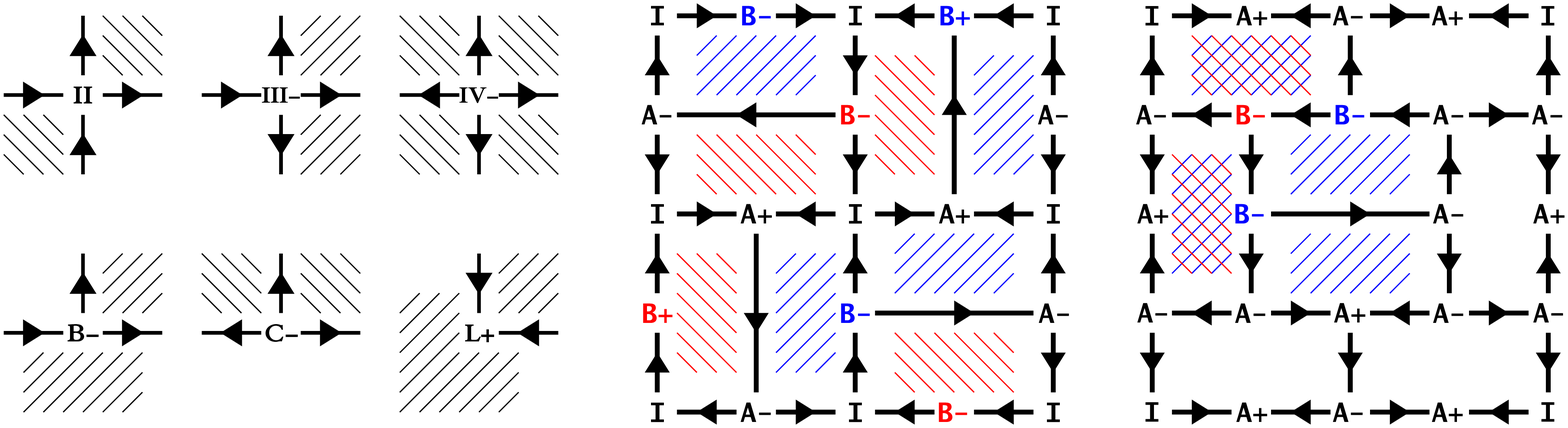}
\caption{
(Color online)
Left: unhappy vertices are shown in isolation, with hatching on defected adjacent loops. Right: Portions of the shakti and Santa Fe lattices. Hatching again indicates frustration, with the color corresponding to the defect providing the frustration.
 \label{Fig:defLoops}
 }
\end{figure}

It is now time to offer a brief analysis of some of the vertex-frustrated lattices we have presented.

Guided by our previous considerations we might ask if it is sufficient to consider only minimal loops when looking for the structure of the ground state in a vertex-frustrated  system. Ideally, the ground state should have a single low-energy unhappy vertex on each frustrated minimal loop  and no unhappy vertices on any non-frustrated loops. 
If such a configuration does exist, it must be a ground state since  it has the minimum number of unhappy vertices and each unhappy vertex is of the ``cheapest'' kind. Put another way, the ground state should accommodate the vertex-frustration in every minimal loop with the lowest possible energy cost.

How can we determine if a given allocation of unhappy vertices accommodates frustration with a minimum energy expenditure?
Figure~\ref{Fig:defLoops} helps answer this question by tabulating all defect types along with which adjacent loops are frustrated.
Note that, except for Type IV, all unhappy vertices provide frustration for only two loops. This means that Type Bs are the most efficient, in terms of energy cost per frustrated loop, so we expect ground states of all vertex-frustrated lattices to favor Type Bs when possible.

\begin{figure}
\includegraphics*[scale=0.4]{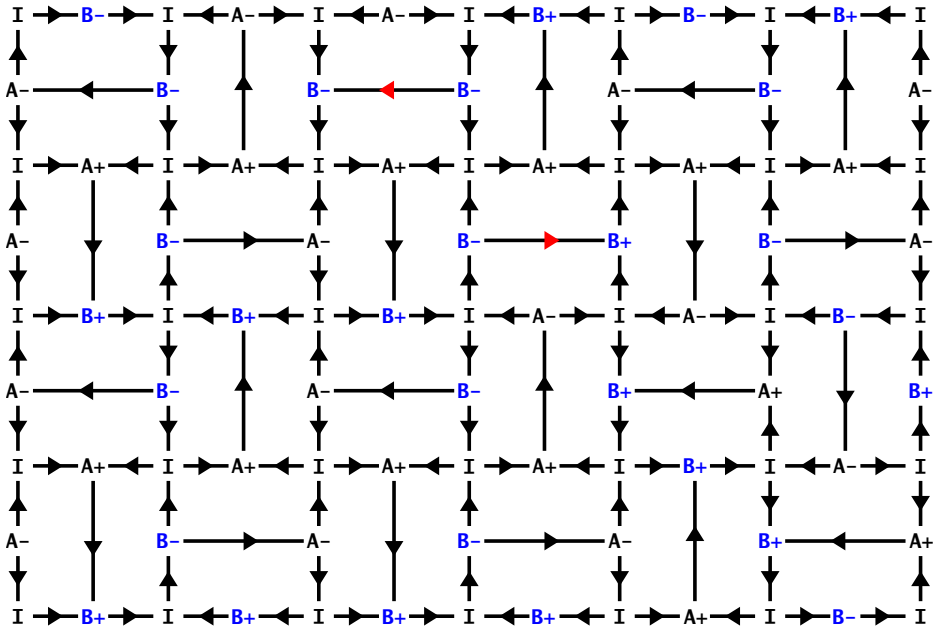}
\caption{
(Color online)
The extensively degenerate ground state of the shakti lattice: each square plaquette must have 2 Type B unhappy vertices (blue) placed somewhere on the 4 available perimeter vertices, consistent with the neighboring squares. Once again, the red spins may be flipped without leaving the ground state, but any other single-spin flip creates new unhappy vertices.
 \label{Fig:checkerBiGS}
 }
\end{figure}

\subsection{Shakti}

The shakti lattice  provides an example in which the ground state does follow from our intuitive reasoning.
Following the above comments, since Type Bs have the lowest energy cost per frustrated loop, one might try to satisfy the vertex-frustration using only Type B unhappy vertices. Figure~\ref{Fig:checkerBiGS} shows that every Type B defect provides frustration to two adjacent loops (the maximum possible without resorting to Type IVs), and every minimal loop receives frustration from only one defect. In a sense this configuration has ``maximal sharing" of unhappy vertices, which illustrates the rule hinted at earlier: if a configuration exists such that every frustrated minimal loop receives frustration from one and only one Type B, it is a ground state. Since each square plaquette contains two frustrated (rectangular) loops, there must be two Type Bs on its perimeter, but they can be located on any of the four sides (as long as they are consistent with neighboring plaquettes).

On the basis of these considerations,  the residual entropy of the shakti lattice can be computed exactly,  and its ground state can be mapped in a thermally disordered  state of the F model, as we will report elsewhere [Chern2012]. As the shakti lattice is \textit{maximally} frustrated, i.e., every minimal loop is frustrated it does not suffer from the complications of induced frustrations.

\subsection{Pinwheel}

The ground state of the pinwheel lattice of Fig.~\ref{Fig:TopFr} is also nontrivial and extensively degenerate.  
First notice that the square minimal loops are not vertex-frustrated, while the T-shaped loops are. Each frustrated loop must host at least one, and possibly as many as three, Type B vertices. This represents maximal sharing of unhappy vertices, as in the shakti: every Type B frustrates two loops, and each loop receives frustration from only one defect. Note that in the ground state each T-loop has only three sites which can host a Type B: the other three-island sites would frustrate the square loops, if occupied. This would cost energy while producing unneeded frustration, so such a configuration cannot be a ground state. 

Although not maximally frustrated, the pinwheel lattice has no induced frustration on nonfrustrated loops. Like the shakti lattice, the pinwheel lattice has a  nontrivial ground state. While the structure of its ground state is clear, an exact computation of its entropy still eludes us.

\subsection{Staggered brickwork}

\begin{figure}
\includegraphics*[scale=0.34]{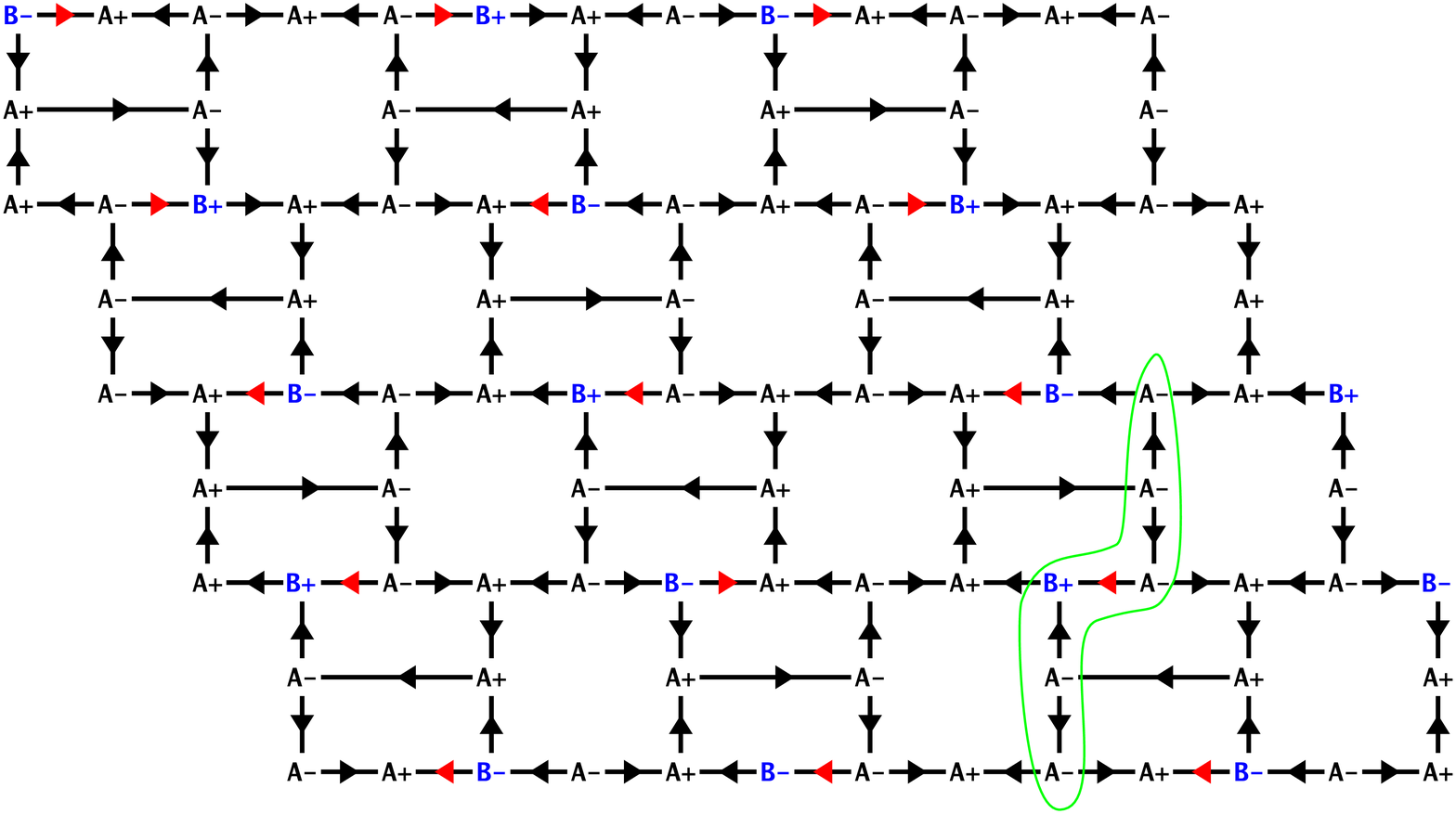}
\caption{
(Color online)
In this lattice, the ground state configuration has extensive degeneracy: each red spin can flip independently of its neighbors at no energy cost. This is equivalent to placing a Type B defect on either vertex adjacent to a red spin, which provides frustration for two adjacent loops. The green circled vertices demonstrate the impossibility of monopole crystallization, discussed in Section~\ref{Sec:Monopoles}. Since the placement in each unit cell is independent of every other unit cell, the degeneracy of the ground state is trivial.
 \label{Fig:anisoGS}
 }
\end{figure}

In contrast with the previous examples, the staggered brickwork lattice of Figure~\ref{Fig:anisoGS} provides an example of {\it trivial degeneracy}: a degeneracy that is completely local and can be counted as $2^{N_c}$, where $N_c$ is the number of independent local choices one can make.

The unit cell of this lattice contains two vertex-frustrated minimal loops and one nonfrustrated loop (the rectangles and square, respectively). Similarly to the shakti lattice, if there exists a configuration with a single Type B providing frustration for two rectangles simultaneously (and no frustration for any square loops), it must be the ground state. 
Figure~\ref{Fig:anisoGS} demonstrates such a configuration, in which each unit cell has a single defect which can be placed on one of two vertices. Note that this collection of configurations exhausts the ground state: it is impossible to accommodate frustration in two loops by placing a Type B anywhere except adjacent to a red spin, and any other type of defect has a higher energy cost for the same frustration. In contrast with the shakti and pinwheel lattices, the choice in one unit cell in completely independent of the choice in neighboring unit cells. Thus the degeneracy is simply \(2^N\) where \(N\) is the number of unit cells, so the entropy density per island is immediately
\(s=(1/9)\ln{2}\)

\subsection{Santa Fe}

As an example of a  nontrivial ground state with induced frustration (see Section~\ref{Sec:FrustInduc}), we return to the Santa Fe lattice of Figure~\ref{Fig:Tammies}.  In each square plaquette (with Type I vertices at the corners) the two rectangles at the center are the only frustrated minimal loops, while the other six minimal loops are nonfrustrated.

This ground state is the most difficult to derive. First note that a single Type B cannot frustrate two frustrated loops. Referring again to Figure~\ref{Fig:defLoops}, this is because a Type B placed on any of the eight sites surrounding the two frustrated loops must frustrate one of the six nonfrustrated loops. Even \textit{two} Type B vertices cannot frustrate the two loops. If such a configuration were possible, they would have to both provide frustration for the \textit{same} nonfrustrated loop, but some inspection shows it is then impossible for them to reach the two frustrated loops. So we are led to consider three Type B vertices to provide frustration for two frustrated loops. In the dumbbell model, three Type B vertices is still less expensive in energy than a single Type C (but this may change in different models for the energetics, as discussed in the appendix). Note also that the four-island vertices are too far from the frustrated loops to possibly participate with a lower energy cost than three Type B vertices.

In Figure~\ref{Fig:Tammies}, the second plaquette from the left in the bottom row provides the prototype. Here three Type B vertices placed around a corner frustrate both frustrated loops in the plaquette while canceling out the frustration on the nonfrustrated minimal loops. But as the rest of the figure shows, frustrated loops in neighboring plaquettes can be linked by three Type Bs for the same total energy cost. There is clearly an enormous number of ways to connect plaquettes together to satisfy the frustration, making the ground state, and the residual entropy,  nontrivial.

Despite this complexity, the ground state of the Santa Fe lattice can be mapped to a thermally disordered state in a vertex model, like we saw for the shakti lattice. In this case the appropriate mapping is to the eight-vertex model, as we will demonstrate elsewhere.

\subsection{Tetris}

\begin{figure}
\includegraphics*[scale=0.36]{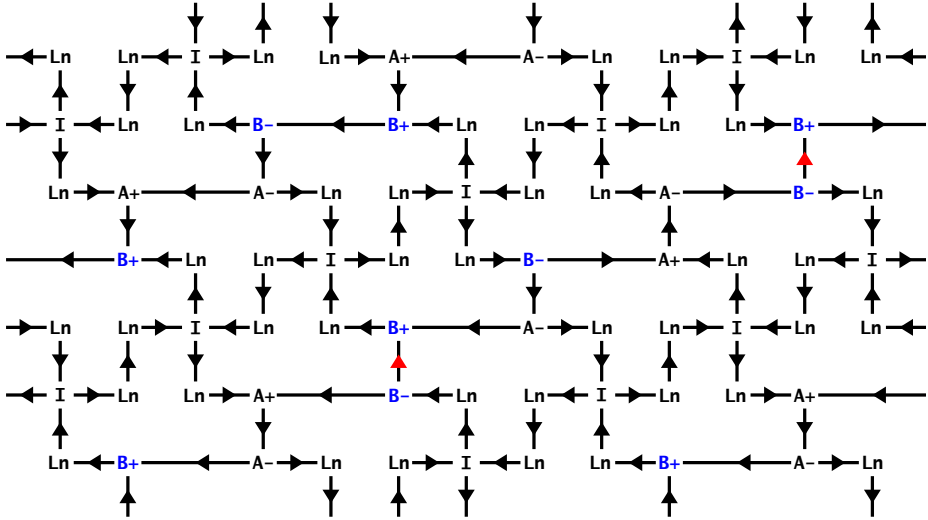}
\caption{
(Color online)
Ground state of the tetris lattice. As before, unhappy vertices are colored blue and red spins can flip while remaining in the ground state. The ground state of this lattice reduces to a sequence of one-dimensional spin ``staircases" running diagonally and connecting Type A and B vertices. The spins on this staircase may either be ordered (as in the upper left) or disordered (as on the main diagonal).
 \label{Fig:doubleT}
 }
\end{figure}

Like the shakti, the tetris lattice (Fig~\ref{Fig:doubleT})  is maximally frustrated, and the configuration shown is a ground state thanks to similar sharing of unhappy vertices. Like the staggered brickwork, it has lower rotational symmetry than many other lattices we have presented--but unlike the staggered brickwork, its degeneracy does not reduce to an independent choice in each unit cell.  Rather interestingly, its ground state is a sliding phase, which decomposes  into parallel  bands of Type A and B vertices (we call these staircases), separated from each other by parallel bands of Type I and Ln vertices (we call these backbones)

In the ground state, the Type I/Ln in the backbone are completely  assigned and ordered. Once a single spin is chosen, the configuration for the entire stripe is determined (including spins on both sides of each Ln vertex). This reduces the lattice to a series of truly one-dimensional problems  as the only remaining free spins form a one-dimensional staircase connecting Type A and B vertices. Depending on the spin choice in Type I/Ln backbones, there are two possible boundary conditions for the Type A/B staircase imposed by the Ln vertices.

The first possibility is shown in the upper left of Figure~\ref{Fig:doubleT} where the (horizontal) boundary spins  from the Ln vertices of the two backbones are parallel. This forces the staircase to form a continuous parallel chain of spins since any other choice would either produce Type C vertices or doubly-frustrate a minimal loop. On the steps of the staircase we have then alternating couples of Type A and Type B vertices. Note that, without changing the boundary conditions one can flip all the spins of the staircase and exchange position for Type A and Type B vertices.

The second possibility is shown in the main diagonal of Figure~\ref{Fig:doubleT} and leads to extensive entropy for the lattice. Here the Ln boundary spins are antiparallel on each horizontal row, which allows considerable freedom in allocating defects. This leaves the staircase disordered with entropy proportional to its length, and since each staircase can be chosen independently, the entire two-dimensional lattice possesses extensive entropy. 
In this second case the Type B vertices posses a certain degree of mobility along the one-dimensional staircase.

\begin{figure}
\includegraphics*[scale=0.40]{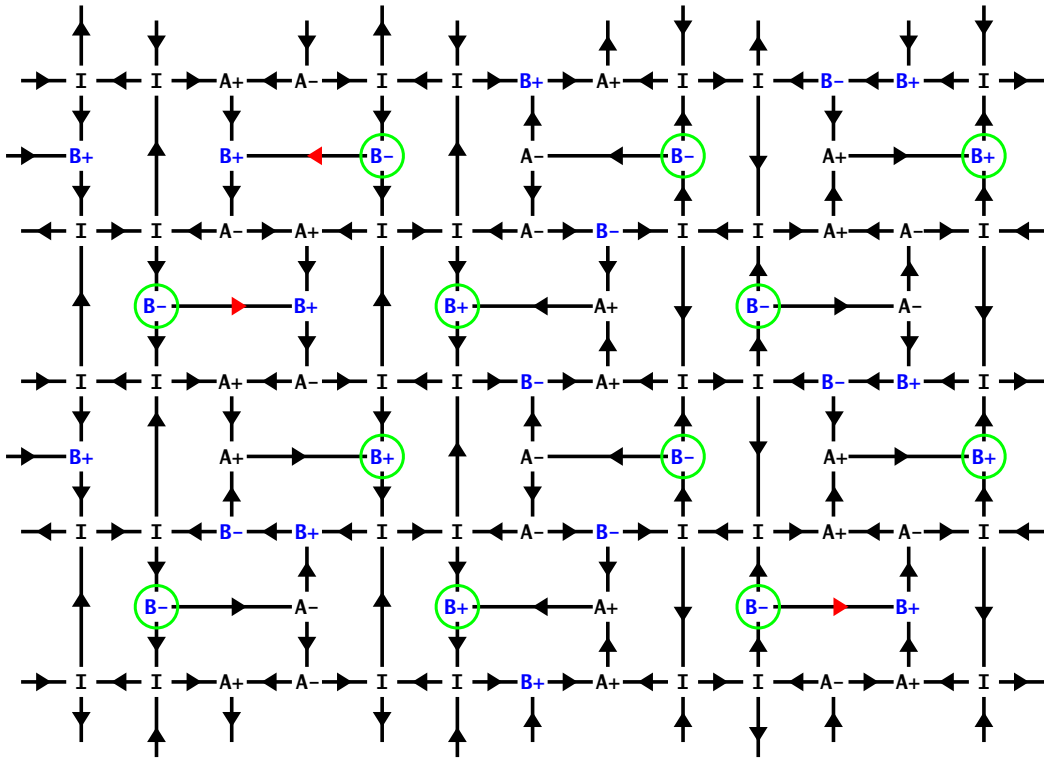}
\caption{
(Color online)
Ground state of the staggered shakti lattice. As above, red spins can flip without energy cost. Similar to the tetris, the ground state breaks into a sequence of noninteracting stripes separated by chains of Type I vertices. It is unusual among the lattices studied in that the (green) circled vertices are \textit{forced} to host an unhappy defect. Despite the stringent restrictions on allocating unhappy vertices (discussed in the text), it nonetheless retains extensive entropy.
 \label{Fig:noMonopoleCryst}
 }
\end{figure}

The final example we present is the staggered shakti lattice in Figure~\ref{Fig:noMonopoleCryst}. Although visually similar to the shakti and staggered brickwork, its ground state more closely resembles that of the tetris lattice. Even though the lattice is maximally frustrated, there is considerably less freedom in allocating defects than in the other maximally frustrated lattices we have presented (the shakti and tetris). We will comment more on this in Section~\ref{Sec:Monopoles}. Note especially the minimal loops with four Type I vertices and only one three-island vertex (circled green in Figure~\ref{Fig:noMonopoleCryst}), which is forced to carry an unhappy Type B vertex in every instance.

The Type I vertices form continuous vertical chains that isolate the bands of Type A/B vertices from each other. As in the tetris, the spin choices on neighboring Type I chains provide 2 possible boundary conditions for the A/B stripes.

The first possibility is depicted in the middle column of Figure~\ref{Fig:noMonopoleCryst}. Here the horizontal boundary spins are anti-parallel, forcing the presence of a Type A and a Type B on each horizontal row connecting two Type Is. But there are also the required circled Type Bs. After a single arbitrary spin choice, no more choice remains in the stripe. To avoid Type C vertices or double-frustration of loops, every spin is determined; the degeneracy of the stripe is two, corresponding to the arbitrary first choice.

Boundary condition number two is shown in the left and right columns of Figure~\ref{Fig:noMonopoleCryst}. 
Since the horizontal boundary spins are parallel, this allows the choice of two Type As or two Type Bs on the horizontal rows connecting Type Is. To be consistent with the forced (circled) Type Bs, two possible tilings are allowed. If Type As are chosen in two consecutive rows, a second Type B must be placed opposite the forced Type B. If Type As are chosen in one row and Type Bs in the next, all spins are entirely determined from the Type A vertices. The two tilings can mix and alternate arbitrarily as Figure~\ref{Fig:noMonopoleCryst} suggests, so the stripe is disordered and provides the entire lattice with extensive entropy.


The reader may observe that all the frustrated lattices presented thus far have ground states consisting of Type I, Type Ln, Type A, and Type B vertices. None of the higher energy unhappy vertices from Figure~\ref{Fig:VertEn} have appeared. This is as we speculated at the beginning of Section~\ref{Sec:Static}, and one may wonder if this is generally true. 
Figure~\ref{Fig:nonObGS} in the appendix shows that, in general, the answer is no although, from the lattices we have studied, this example appears to be the exception rather than the rule.

\begin{figure}
\includegraphics*[scale=0.32]{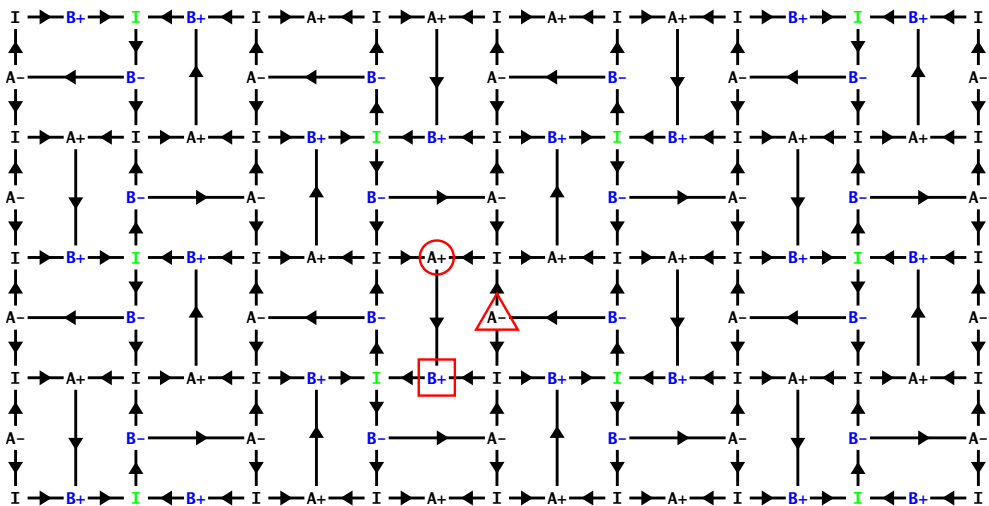}
\caption{
(Color online)
The shakti lattice from Figure~\ref{Fig:checkerBiGS} in its monopole crystallized state. Groups of 4 unhappy vertices (blue) cluster around a Type I vertex (green) of opposite spin orientation from the background. 
Note that the circled and boxed vertices are \textit{topologically} nearest neighbors but \textit{geometrically} third-nearest neighbors, while the vertex in the triangle is a second-nearest neighbor in both senses to the circled and boxed vertices. Even though this configuration places like magnetic charges on topological nearest-neighbors, it still minimizes Coulomb repulsion between vertices.
This monopole crystallized configuration also exhibits spontaneous isotropy breaking, as discussed in the text.
 \label{Fig:checkerBiCr}
 }
\end{figure}

\section{Monopole Crystallization and Smectic Phases}
\label{Sec:Monopoles}

 We have shown several examples of vertex frustrated ASI lattices with an extensively degenerate, disordered ground state. In the only current realization of degenerate ASI, the heavily studied hexagonal lattice~\cite{Qi2008,Lammert2010,Moller2009}, this degeneracy is lifted in theory by longer-range interactions~\cite{Moller2009, Chern2011}. The lattice can then enter a ``crystallized monopoles" state in which neighboring vertices posses opposite magnetic charges; note that this state is still disordered and contains extensive entropy, though much less than the pseudo-ice manifold~\cite{Lammert2010,Moller2009, Chern2011}. 
Although this state as not yet been observed experimentally in ASI, signatures of long range dipolar interactions have been reported~\cite{Rougemaille2011}, as well as direct visualization of magnetic monopoles in polarized honeycomb lattice, following magnetic reversal~\cite{Ladak2010, Mengotti2010}
 
A natural question is if a similar ``monopole crystallization" can occur in topologically frustrated lattices. A complete study must include the effects of long range interaction and goes beyond our aim. Here we can obtain some insight by looking  for sub-manifolds of the ground state in which nearest neighboring vertices harbor magnetic charges of opposite sign.  

\begin{figure}
\includegraphics*[scale=0.30]{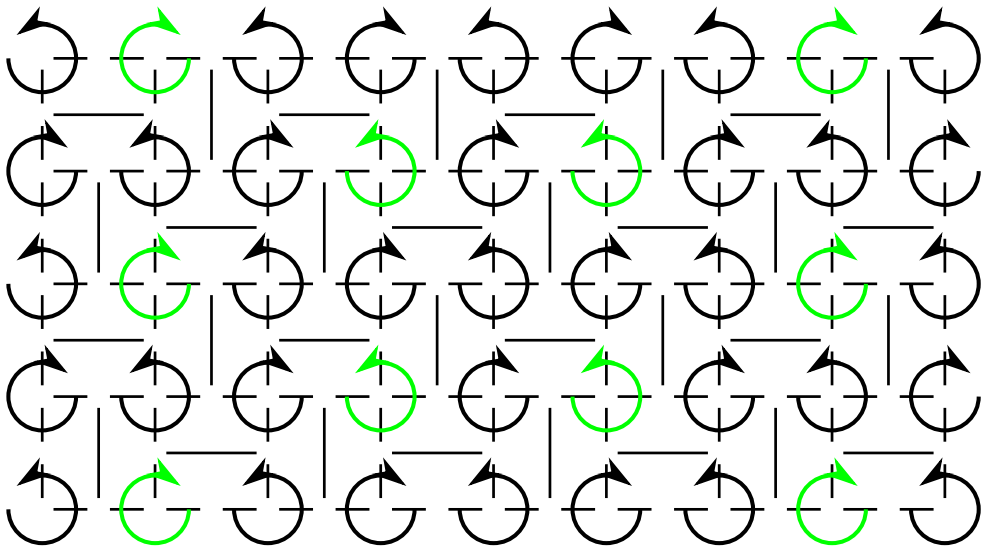}
\caption{
(Color online)
The monopole crystallized configuration from Figure~\ref{Fig:checkerBiCr}. The lattice is represented as two sublattices of Type I vertices with opposite chiralities. Green arrows represent ``occupied" sites, corresponding to the green Type I vertices surrounded by Type B defect clusters in Figure~\ref{Fig:checkerBiCr}. If only sites of one chirality are occupied, the spin configuration remains isotropic and ordered, but occupying sites of different chirality defines ordered and disordered directions (as drawn, vertical and horizontal, respectively).
 \label{Fig:pinwheelSites}
 }
\end{figure}

For the shakti lattice such monopole crystallization is possible, as shown in Figure~\ref{Fig:checkerBiCr}. In this lattice, since three-island vertices have uncharged Type I vertices as nearest neighbors, this configuration places opposite charges on second nearest neighbors. 

Interestingly,  the shakti lattice presents spontaneous symmetry breaking into a smectic phase, even though the underlying lattice has no preferred direction. Notice that, as drawn, every three-island vertex with a vertical line passing through it has a negative monopole charge, and conversely, every three-island vertex on a horizontal line has a positive charge. This clearly reduces rotational symmetry from four-fold to two-fold.

A peculiar consequence of this is that the spins are perfectly ordered in one direction (vertical, as drawn) and disordered in the orthogonal direction. This occurs with the choice of where defect clusters are placed. We can associate a chirality with each defect cluster according to the direction in which the four nearest long islands curl. 
Figure~\ref{Fig:pinwheelSites} illustrates the configuration from Figure~\ref{Fig:checkerBiCr} with this notation. The lattice may be viewed as two interleaved lattices of left- and right-handed Type I vertices.

In principle, one could build a unique, perfectly ordered state by placing all defect clusters around Type I's of the same handedness. Far more entropically favorable, however, would be a mixture. Note that each defect cluster serves a supercell of four plaquettes. The ordered/disordered directions are determined by the occupation of two adjacent supercells of opposite handedness. Then there is no choice of how to occupy adjacent supercells without violating the ground state rules. Rows of clusters of a particular handedness may only be extended to infinity in one direction, defining the ordered axis, while the perpendicular direction remains disordered.

\begin{figure}
\includegraphics*[scale=0.30]{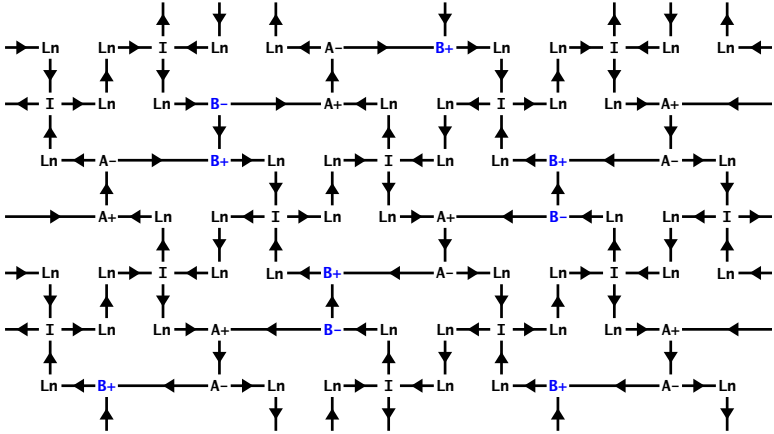}
\caption{
(Color online)
The monopole crystallized configuration for the tetris lattice from Figure~\ref{Fig:doubleT}. Each Type A/B staircase has two possible configurations, obtained by flipping all horizontal spins on a staircase, so the lattice is infinitely degenerate but without extensive entropy, like the shakti above.
 \label{Fig:doubleTCrystal}
 }
\end{figure}

Like the shakti, the tetris lattice from Figure~\ref{Fig:doubleT} also exhibits a monopole crystalline state, shown in Figure~\ref{Fig:doubleTCrystal}. Like the shakti's, although less surprisingly so, monopole state, its crystallized ground state  is infinitely degenerate but extensively only in one direction. This is because, as noted in the discussion of Figure~\ref{Fig:doubleT}, the lattice  decomposes into a series of one-dimensional staircases, separated by backbones of Type I and Ln vertices. in the crystallized state not only the backbones, but also the staircase are ordered. While the backbones are assigned, the crystallized staircases can flip all their spins independently from each other and from the backbones. Therefore, the monopole crystalline state of the tetris  is perfectly ordered parallel to the stripes and disordered in the orthogonal direction.

None of the other lattices presented in this work allow for monopole crystallized states under any reasonable definition. Intuitively, the crystalline condition dramatically reduces the entropy by placing a powerful extra constraint on how unhappy vertices can be arranged. Such a configuration is only possible if the original vertex ground state has sufficient freedom for the placement of unhappy vertices.

As an example, we show why this is so for the staggered brickwork lattice in Figure~\ref{Fig:anisoGS}. In this case, the vertical chains of three like-charged Type A vertices cannot be removed while remaining in the energetic ground state. This can be seen by considering the string of six vertices highlighted in Figure~\ref{Fig:anisoGS}. All must be Type A except for the middle two, only one of which may be a Type B. Whichever one is chosen, a string of three Type As remains, all with the same charge (choosing the charge of a Type A is equivalent to choosing one spin, and determines the charge on all neighboring Type As). 

The impossibility of crystallization for the Santa Fe and pinwheel lattices follows from a similar argument. Notice that these lattices are not maximally frustrated, whereas the shakti and tetris are. One might wonder if there is a connection between maximal frustration and monopole crystallization. Figure~\ref{Fig:noMonopoleCryst} shows this is in general not the case. This lattice, which is maximally frustrated, does not admit a monopole crystalline configuration. A maximally frustrated lattice usually has enormous freedom in allocating defects, judging from our earlier examples. But in this case, there exist loops with four Type I vertices and only a single three-island vertex, which therefore must hold an unhappy vertex in the ground state. With some experimentation, one can see it is then impossible to place opposite charges on nearest neighbor three-island vertices.

This shows that maximal frustration does not imply the possibility of monopole crystallization, but does nonmaximal frustration preclude crystallization? We strongly suspect this is might be the case but have been unable to find a proof or a counterexample. Nonfrustrated minimal loops (usually) must not hold unhappy vertices. This is a strong restriction, as is the monopole crystallization condition, so unless the original ground state had large degeneracy, it may be impossible to find any configuration that can satisfy both restrictions.

\section{Conclusion and perspectives}
\label{Sec:Dynamics}

We have introduced a new source of frustration in vertex models designed to fabricate artificial spin ice of desired residual entropy, and we have shown specific lattices that manifest such frustration. We have discussed their degeneracy, trivial and not, and hinted at the possible effect of long range interactions in eliciting  emergent symmetry breaking.

Current advances in thermalization of ASI~\cite{Morgan2010,Kapaklis2012} should allow for their  topologically vertex-frustrated ground states to be revealed in experimental settings. These new kind of lattices should provide more freedom in ASI design, but also might show their utility in the study of field driven magnetically charged excitations. The tetris lattice is an example of  this motion can be constrained to a direction. Bramwell and collaborators have measured long-lived magnetic monopole currents in natural spin ice~\cite{Giblin2011}. Yet, while extremely low temperature conditions and tiny magnetic fields greatly limit advances and potential applications of magnetricity in spin ice, ASI promises to be a more docile medium, especially as a more dynamical ASI, with lower coercive fields and Curie temperatures~\cite{Kapaklis2012}, are being developed.
Since the interaction and screening properties of magnetic monopoles depend upon the geometry of the array, vertex frustration can be exploited to tailor design those properties. 

In many vertex frustrated geometries, unhappy vertices can change their position. Yet their motion is in general not free, but rather crosses kinetic barriers corresponding to the collective flip of many spins which entails creation of higher energy vertices. At zero or low temperature and low field it should be possible to think of these processes in terms of creation, propagation and annihilation of excited vertices, as  the system is  driven to explore the   degenerate manifold of its ground state.

Finally, we have reasoned here in terms of island, but most of our concepts applies to contiguous lattices which might be employed for experiments in magneto transport~\cite{Branford2012}. 

\section{Credits and Acknowledgments}

In this work, Muir Morrison, Tammie Nelson, and Cristiano Nisoli contributed the design of lattices. Muir Morrison, with help from Cristiano Nisoli alanyzed the general properties of vertex-frustration. Muir Morrison provided the study of all the ground states proposed here. Cristiano Nisoli proposed the idea of extensive degeneracy from non-degenerate vertices through vertex-frustration. We are grateful to A. Libal, G.-W. Chern, C. Reichhardt, P. Lammert, V. Crespi for useful discussions. This work was carried out under the auspices of the US Department of Energy at LANL under contract no. DEAC52-06NA253962, and speci�cally LDRD grant no. 20120516ER.

\section{Appendix}

Obviously the dumbbell model which we chose for definiteness is only one way to assign vertex energies.  The specific choice of the energetics has to do with parameters in the physical realization of the lattice, e.g. the ratio between islands and lattice constant. Another possibility is to assign pairwise interaction energies by treating each island as a point dipole.  In general this does not change the energetic hierarchy of single vertices in each group (I-IV and A-C). In fact \textit{any} choice  that  assigns a stronger interaction to perpendicular spins compared with collinear spins replicates the same vertex hierarchy in each group of vertices. Nonetheless a different choice might change the relative ratio of energy difference between the two groups, and allow for more, or less, degeneracy.

\begin{figure}
\includegraphics*[scale=0.3]{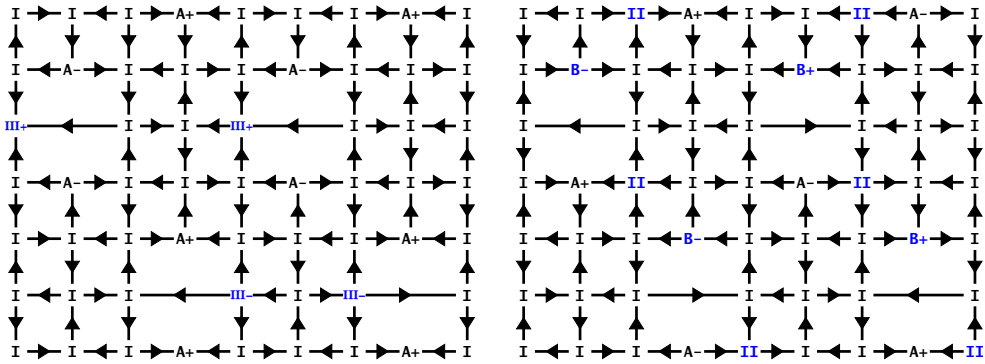}
\caption{
(Color online)
In this lattice, the configuration with the minimum number of  unhappy vertices (blue) is shown at left. However, this is not the ground state, shown at right, in which even unfrustrated minimal loops are forced to carry unhappy vertices. We can show this configuration is a ground state as follows. Referring to Figure~\ref{Fig:defLoops}, only a Type III defect can simultaneously frustrate two neighboring rectangular loops (as in the configuration at left). The next best we can achieve is using two unhappy vertices to frustrate two rectangular loops. This cannot be done with two Type Bs (the lowest energy choice) since the three-island vertices are third-nearest neighbors. A Type II and a Type B, as shown at right, is thus the lowest energy configuration. On the other hand, with a different choice of energetics that treats the islands as point dipoles, the configuration on the left represents the ground state, which has therefore trivial degeneracy. 
 \label{Fig:nonObGS}
 }
\end{figure}

Of the ground states presented whose who contain  only unhappy vertices of  Type Bs, like the Shakti lattice, are insensitive to the choice of energy model, as long as it assigns a stronger interaction to perpendicular spins compared with collinear spins. Complication occurs in lattices such as shown in Figures~\ref{Fig:Tammies}~and~\ref{Fig:nonObGS}: the ground states shown are calculated in the dumbbell model, in which the energy cost of three Type Bs, or two Type Bs and one Type II, is significantly less than the cost of a single Type III. In the dipole model, the opposite is true by a narrow margin.

\begin{figure}
\includegraphics*[scale=0.38]{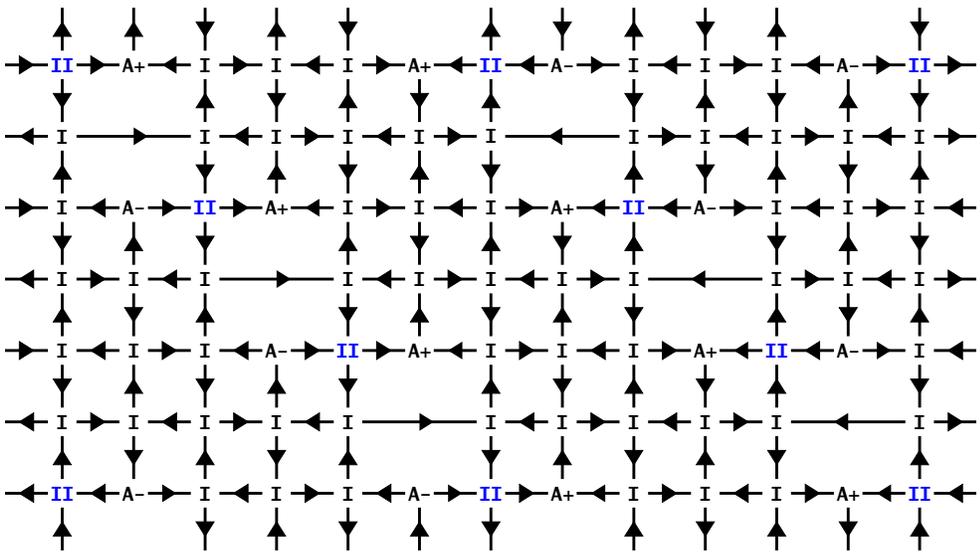}
\caption{
(Color online)
This lattice is frustrated, yet the ground state is nondegenerate. Each Type II is shared by two frustrated loops, and no lower energy configuration is possible using Type B unhappy vertices, or a combination of Type B and Type II unhappy vertices.
 \label{Fig:nonDegenFrust}
 }
\end{figure}

In this appendix we also provide a counterexample for a lattice which is vertex-frustrated, but has a unique ground state, in Fig~\ref{Fig:nonDegenFrust}.


\newpage

\bibliography{library}

\end{document}